\documentclass[final,12pt]{elsarticle}

\usepackage{lineno}
\usepackage{amsmath,amsfonts,amsthm,amssymb,textcomp}
\usepackage{graphicx}
\usepackage{xcolor}
\usepackage{booktabs}
\usepackage{bm}
\usepackage{tikz}
\usepackage{rotating}
\usepackage{comment}
\usepackage{soul}
\usepackage{csquotes}
\usepackage{wrapfig}
\usepackage{float}
\usepackage[export]{adjustbox}
\usepackage{subcaption}
\usepackage{pdflscape}
\usepackage{tablefootnote}
\usepackage{fix-cm}
\usepackage{hyperref}

\modulolinenumbers[5]

\graphicspath{{./images/}}

\journal{Aerospace Science \& Technology}

\biboptions{sort&compress}

\begin{document}

\title{Generalized Formulation to Predict Rossiter Modes for Subsonic to Hypersonic Flow}
\author[ucaero,df]{Jeremy Redding}
\author[arl]{Luis Bravo}

\author[ucaero,df]{Prashant Khare\corref{cor1}}
\cortext[cor1]{Corresponding author email: Prashant.Khare@uc.edu}
\address[ucaero]{Department of Aerospace Engineering, University of Cincinnati, Cincinnati, OH 45221-0070}
\address[df]{Hypersonics Laboratory, Digital Futures, University of Cincinnati, Cincinnati, OH 45206}
\address[arl]{DEVCOM Army Research Laboratory, Army Research Directorate, Aberdeen Providing Ground, MD, 21005}

\date{\today}

\begin{abstract}
This paper describes the development of a generalized, physics-based theoretical model for estimating Rossiter modes for flow over rectangular cavities for regimes ranging from subsonic to hypersonic without requiring a priori knowledge of the detailed cavity-flow dynamics. This model addresses the issues with the Heller Bliss model, which despite incorporating compressibility effects, shows increasing deviations from 
experimentally measured Rossiter modes above Mach 3. The proposed model exploits the fact that the acoustic propagation speed governing the Rossiter feedback mechanism depends on 
the excitation and relaxation of internal energy modes. To account for these effects, an effective temperature is constructed from estimated post-shock conditions and used to evaluate an effective speed of sound, which is then used in the Rossiter relation to calculate the Strouhal number. The Strouhal numbers predicted by the model are compared with available experimental and high-resolution two-dimensional numerical simulations (conducted as a part of this study) from Mach 2 to Mach 9; our results agree very well with these datasets, thus 
supporting the conclusion that thermodynamic energy-mode participation and relaxation can alter the acoustic feedback timescale governing cavity oscillations. The formulation therefore provides a physics-based estimate of Rossiter-mode frequencies without requiring prior CFD or experimental characterization of the detailed cavity-flow dynamics.

\end{abstract}

\maketitle


\section{\label{sec:introduction}Introduction}

When flow passes over a rectangular cavity, the incoming flow separates at the leading edge and forms a shear layer over the cavity. In the classical Rossiter interpretation, the resulting oscillatory cycle consists of shear-layer disturbance formation, downstream convection, and impingement near the trailing wall. The impingement generates pressure waves that propagate upstream and excite new disturbances near the leading edge, thereby closing the feedback loop that sustains the oscillation.

The pressure field within the cavity is inherently unsteady and may further amplify instabilities at the leading edge. These instabilities have attracted sustained research interest because periodic pressure fluctuations can excite structural modes and thereby generate potentially hazardous flight conditions. In his seminal study, \citet{rossiter1964wind} investigated these pressure-driven instabilities and identified several key physical mechanisms that underpin the present work. This work showed that shallow cavities with $L/D \geq 10$ are strongly influenced by the dynamics of the boundary layer, since the thickness of the boundary layer is comparable to the depth of the cavity. Under these conditions, the random component of the pressure fluctuations becomes comparatively more significant, leading to a broader frequency spectrum without a single clearly dominant peak, although two peaks were observed in some cases. In contrast, for deeper cavities with $L/D=1 \text{ or } 2$, which are the focus of the present study, the periodic component of the pressure fluctuations predominates, while the random component associated with boundary-layer thickness and instability is minimal. Rossiter further noted that, for a cavity of fixed geometry, increasing the freestream Mach number amplifies the periodic component, rendering it more distinct.

In the same work, \citet{rossiter1964wind} proposed an expression for predicting the periodic component of cavity pressure fluctuations in terms of a cavity-based Strouhal number. As shown below, this expression is derived from thin-sheet oscillation theory, with coefficients determined from experimental data:
\begin{equation}
\label{eq:rossiter_org}
    St =
    \frac{n-\alpha}
    {1/\kappa_v+M_{\infty}}.
\end{equation}

\noindent
In this equation, $n$ is the mode number, $M_{\infty}$ is the freestream Mach number, $\alpha$ is the phase-delay parameter, and $\kappa_v$ is the ratio of the characteristic shear-layer convection velocity to the freestream velocity. Although multiple Rossiter modes are considered in the present analysis, it is useful to note that Rossiter's original experiments identified the second mode as the most strongly amplified mode under the conditions examined. Subsequent studies have likewise frequently observed strong amplification of the second Rossiter mode, although the dominant mode can vary with cavity geometry and flow conditions \citep{rossiter1964wind,heller1975physical,smith2024history,sun2020resolvent,colonius1999numerical,liu2021unsteady}. Therefore, comparisons with the second mode are emphasized where appropriate. Similarly, consistent with previous studies, including the original study of \citet{rossiter1964wind}, our analysis is based on pressure measurements at $x/L=0.9$ on the lower wall of the cavity. Equation~\ref{eq:rossiter_org} was developed from Rossiter's subsonic and transonic experiments and does not include the later compressibility correction introduced for higher-speed flows.

To account for compressibility, \citet{heller1975physical} modified Eq.~\ref{eq:rossiter_org} using Crocco's theorem, as shown below:
\begin{equation}
\label{eq:heller}
    St =
    \frac{n-\alpha}
    {M_{\infty}/\sqrt{1+\frac{\gamma-1}{2}M_{\infty}^2}
    +1/\kappa_v}.
\end{equation}

\noindent
This form has been shown in the literature to accurately predict Strouhal numbers for flows up to approximately Mach 3 \citep{heller1975physical}. More recently, \citet{dechant2019cavity} proposed a depth-sensitive extension of the Rossiter formulation that accounts for changes in cavity depth. Such a correction becomes especially useful as a cavity approaches the shallow-cavity regime, where boundary-layer instabilities and turbulent inflow can have a more significant influence on the pressure oscillations. This equation contains fewer empirical coefficients and reasonably predicts Rossiter modes up to approximately Mach 3. However, recent studies of hypersonic flow over rectangular cavities \citep{redding2023thermochemical,redding2025vortex,benitez2026large} indicate that these models begin to diverge from observed Strouhal numbers for Mach numbers beyond approximately 4.

At higher Mach numbers, experimental evidence also suggests that the character of the cavity oscillation can become increasingly acoustic. For example, \citet{unalmis2001experimental} found that the measured resonance frequencies of a Mach 5 rectangular cavity approached the longitudinal acoustic modes of a closed cavity. Their results suggested that coupling between the shear layer and the cavity acoustic field became weaker at high Mach number, such that a simple closed-box acoustic model could predict the measured frequencies reasonably well. Previous high-Mach-number studies have also questioned whether the strongly coupled shear-layer mechanism associated with the classical interpretation of Rossiter oscillations persists unchanged throughout the entire Mach-number range. This observation is important because it suggests that high-Mach-number cavity behavior may approach an acoustically dominated limiting state rather than simply following an extrapolation of the classical Rossiter relationship.

However, a closed-box acoustic model does not retain the convective component of the Rossiter feedback cycle and therefore does not describe the transition from the classical shear-layer/acoustic feedback mechanism to this increasingly acoustic-dominated behavior. Furthermore, the closed-box formulation used by \citet{unalmis2001experimental} evaluates the longitudinal resonance using the stagnation speed of sound and therefore does not account for changes in the post-shock thermodynamic state or finite-rate internal-energy relaxation. A model intended to span subsonic through hypersonic conditions should therefore retain the essential Rossiter feedback structure while allowing the acoustic timescale to evolve with the thermodynamic state of the flow.

In this manuscript, we present a theoretical analysis aimed at developing such a model for predicting Rossiter modes from subsonic through hypersonic flow regimes. Rather than replacing the Rossiter feedback mechanism with a purely acoustic closed-box representation, the proposed formulation retains the convective and phase-delay components of the Rossiter model while modifying the acoustic timescale through a thermodynamically dependent effective speed of sound. The effective acoustic state is constructed from normal-shock conditions and modal-energy partition, with a physically bounded effective-temperature branch alongside a vibrational included temperature model separated by a switching Mach number determined from the finite-rate vibrational-relaxation response. In this way, the model is intended to provide a reduced-order connection between the classical Rossiter description and the increasingly acoustic-dominated behavior observed at higher Mach numbers.

After formulating the revised model, we assess its predictive accuracy using two-dimensional direct numerical simulations of thermochemical nonequilibrium flow over a rectangular cavity for Mach numbers from 2 to 9. We then further assess the generalizability of the model using available experimental rectangular-cavity frequency data spanning Mach numbers from 2 to 6. It is important to emphasize that the numerical simulations are used solely to evaluate the theoretical model and play no role in determining the effective-temperature formulation or its physical switching criterion. Qualifying measured frequency data for directly comparable open rectangular cavities above Mach 6 were not identified in the present literature survey; consequently, the Mach 7-9 comparisons should be interpreted as numerical model evaluation.

The purpose of this theoretical development is to reasonably predict periodic cavity frequencies and the corresponding Strouhal numbers across a broad range of Mach numbers without \textit{a priori} knowledge of the detailed physicochemical state within the cavity. In particular, the model is intended to retain the simplicity of a reduced-order Rossiter-type formulation while incorporating the thermodynamic changes that increasingly influence the acoustic propagation timescale as flow enthalpy increases.

Before moving forward, we note a few caveats and limitations. First, the use of two-dimensional simulations prevents the direct representation of three-dimensional turbulence production through vortex stretching. The present geometry is nominally two-dimensional, and the simulations reproduce the dominant Strouhal-number behavior reported at lower Mach numbers; nevertheless, three-dimensional effects may influence the detailed shear-layer dynamics and oscillation amplitudes. Second, the incoming boundary-layer thickness is not independently varied in the present study. A finite upstream development length is included, allowing a laminar boundary layer to develop before reaching the cavity, but the sensitivity of the model to boundary-layer thickness is left for future investigation.

The remainder of the manuscript is organized as follows. Section~\ref{sec:theory} details the theoretical development of the model, including the thermodynamic scaling and relaxation-based transition criterion. Section~\ref{sec:test} evaluates the model using the numerical simulations spanning Mach numbers from 2 to 9 and comparisons with available experimental data. The major conclusions of the study are then summarized.

\section{\label{sec:theory}Theoretical Model Development}

The dominant frequencies within a cavity are dependent on the geometry of the cavity, the Mach number in the flow $M_{\infty}$, and the Reynolds number of the flow. Fundamental frequencies in subsonic and transonic rectangular cavities based on \citet{rossiter1964wind} are given by
\begin{equation}
    f_m
    =
    \frac{U_{\infty}}{L}
    \left[
    \frac{n-\alpha}
    {M_{\infty}+1/\kappa_v}
    \right],
    \label{eq:rossiter_original}
\end{equation}
where $n$ is the Rossiter mode number, $\alpha$ is the phase-delay parameter, and $\kappa_v$ represents the ratio between the disturbance-convection velocity and the freestream velocity. As mentioned earlier, this equation was extended by \citet{heller1975physical} to account for compressibility and subsequently by \citet{dechant2019cavity} to account for cavities with different $L/D$.

Although the Heller-Bliss correction extends the Rossiter formulation to supersonic conditions, its original validation database extends only to approximately ($M_\infty = 3$). At higher Mach numbers, experiments by \citet{unalmis2001experimental} at ($M_\infty \approx 5$) demonstrated changes in the coupling between the cavity shear layer and the acoustic field, with measured resonance frequencies approaching the longitudinal acoustic modes of the cavity. These observations do not imply the disappearance of Rossiter-type oscillations, since the fundamental feedback process may still consist of downstream-growing instability waves coupled to upstream-propagating disturbances. Rather, they indicate that the classical Heller-Bliss description may not fully capture the evolving propagation and coupling mechanisms in the hypersonic regime. Consequently, application of the Heller-Bliss formulation beyond its original validation range should be regarded as an extrapolation, while the persistence of downstream instability growth and upstream propagation provides a physical basis for the modified feedback formulation developed here.


Therefore, to develop a model that remains applicable at higher Mach numbers, we return to the original feedback-loop interpretation of the Rossiter equation.

The oscillatory behavior of the cavity feedback loop can be represented as
\begin{equation}
    e^{i\omega t}e^{i\varphi}=1.
    \label{eq:feedback_loop}
\end{equation}
This is the equation for a closed feedback loop. In practice, cavity feedback does not behave as a completely closed system because of the mass outflow occurring near the trailing edge \citep{rockwell1978self}. In spite of this, it has been a useful approximation and is retained here. When the time component of the feedback equation is separated into an acoustic contribution and a shear-layer contribution, the equation becomes
\begin{equation}
    e^{i\omega(\tau_a+\tau_c)}e^{i\varphi}=1,
    \label{eq:feedback_two_times}
\end{equation}
which requires
\begin{equation}
    \omega(\tau_a+\tau_c)+\varphi=2\pi n.
    \label{eq:feedback_phase}
\end{equation}

If the convective time is $\tau_c=L/U_c$, where $L$ is the cavity length and $U_c$ is the convective velocity, Eq.~\eqref{eq:feedback_phase} becomes
\begin{equation}
    \frac{\omega}{2\pi}
    \left(
    \frac{L}{U_c}+\tau_a
    \right)
    +
    \frac{\varphi}{2\pi}
    =
    n.
    \label{eq:feedback_convective}
\end{equation}
Defining
\begin{equation}
    f=\frac{\omega}{2\pi},
    \qquad
    St_L=\frac{fL}{U_{\infty}},
    \qquad
    \alpha=\frac{\varphi}{2\pi},
\end{equation}
gives
\begin{equation}
    St_L
    =
    \frac{n-\alpha}
    {\dfrac{U_{\infty}}{U_c}
    +
    \dfrac{U_{\infty}}{L}\tau_a}.
    \label{eq:strouhalmod}
\end{equation}

Taking
\begin{equation}
    \kappa=\frac{U_c}{U_{\infty}},
\end{equation}
and using the Rossiter assumption $\tau_a=L/a_{\infty}$ gives
\begin{equation}
    St_L
    =
    \frac{n-\alpha}
    {\dfrac{1}{\kappa}+M_{\infty}}.
    \label{eq:rossiter_recovered}
\end{equation}

Since $\alpha$ and $\kappa$ appear to change relatively little in the literature, the acoustic timescale $\tau_a$ is the term that must be assessed most carefully as the flow moves into the hypersonic regime. This is consistent with the reasoning of \citet{heller1975physical}, whose correction modifies the acoustic portion of the original Rossiter feedback model to account for compressibility. In the original Rossiter formulation, the assumption that the acoustic propagation speed in the cavity could be represented by the freestream speed of sound was reasonable for the subsonic and transonic conditions considered. At higher Mach numbers, however, the thermodynamic state inside and immediately above the cavity can differ substantially from the freestream state.

For sufficiently high-enthalpy flows, the acoustic response is also influenced by the excitation of internal molecular energy modes and, at still higher enthalpies, by changes in species composition. Thus, the use of a constant-$\gamma$ perfect-gas speed of sound evaluated only from the freestream temperature does not provide a complete physical description of the acoustic propagation process. From basic gas-dynamic theory \citep{vincenti1966introduction}, the speed of sound of an ideal dissociating gas can be written in chemically frozen and equilibrium limits as
\begin{equation}
    a_f^2
    =
    \frac{RT(4+\beta)(1+\beta)}{3},
    \label{eq:frozen_sound_speed}
\end{equation}
and
\begin{equation}
    a_e^2
    =
    \frac{
    RT\beta(1-\beta)(1+2\tau)
    +(8+3\beta-\beta^3)\tau^2
    }
    {
    \beta(1-\beta)+3(2-\beta)\tau
    },
    \label{eq:equilibrium_sound_speed}
\end{equation}
where $\tau=T/\theta$, $\theta$ is the dissociation temperature of the presumed diatomic molecule, and $\beta$ is the level of dissociation, with $\beta=0$ corresponding to no dissociation and $\beta=1$ corresponding to complete dissociation. These expressions provide motivation for the increasing importance of the thermodynamic state at high enthalpy; they are not used directly in the reduced-order model developed below.

The next logical question is therefore the impact of temperature on Strouhal number. Although the previous models perform well at lower Mach numbers, they do not explicitly incorporate the effect of increasing flow enthalpy. To reiterate, the underlying concept of the present analysis is to estimate the Strouhal number without requiring prior knowledge of the resolved conditions inside the cavity. We therefore propose a physics-based effective-temperature model in which the thermodynamic states used to modify the acoustic timescale are determined from normal-shock conditions, modal-energy partition, and vibrational relaxation.

\subsection{Post-Shock Thermodynamic State}
\label{sec:post_shock_state}

As flow enters a supersonic or hypersonic cavity, the freestream flow passes through rapid compression at the leading edge due to interaction of the incoming boundary layer with the recirculating flow in the cavity. This compression produces an unsteady shock structure. Away from the wall, this structure may be oblique and its instantaneous angle depends on both Mach number and the phase of the cavity oscillation. Due to this unsteady behavior, accurately predicting the shock angle without resolving the cavity flow would require additional calculations or experiments. Instead, because rapid compression occurs near the leading edge, we use a normal shock as a first-order analytical estimate of the thermodynamic state immediately above the cavity shear layer.

For a calorically perfect upstream state, the normal-shock pressure and density ratios are
\begin{equation}
    \frac{p_2}{p_{\infty}}
    =
    1+
    \frac{2\gamma}{\gamma+1}
    \left(
    M_{\infty}^2-1
    \right),
    \label{eq:normal_shock_pressure}
\end{equation}
and
\begin{equation}
    \frac{\rho_2}{\rho_{\infty}}
    =
    \frac{
    (\gamma+1)M_{\infty}^2
    }
    {
    (\gamma-1)M_{\infty}^2+2
    }.
    \label{eq:normal_shock_density}
\end{equation}
The post-shock temperature is then
\begin{equation}
    \frac{T_2}{T_{\infty}}
    =
    \frac{
    p_2/p_{\infty}
    }
    {
    \rho_2/\rho_{\infty}
    },
    \label{eq:normal_shock_temperature}
\end{equation}
and the post-shock Mach number is
\begin{equation}
    M_2^2
    =
    \frac{
    1+\dfrac{\gamma-1}{2}M_{\infty}^2
    }
    {
    \gamma M_{\infty}^2-\dfrac{\gamma-1}{2}
    }.
    \label{eq:normal_shock_M2}
\end{equation}

The temperature $T_2$ represents a simple estimate of the temperature immediately above the cavity and near the shear layer. The gas participating in the cavity feedback process is also influenced by recovery near the shear layer and cavity interior. Therefore, to evaluate the modal energy distribution, we define the post-shock cavity temperature, $T_{cav,2}$ as: 
\begin{equation}
    T_{\mathrm{cav},2}
    =
    T_2
    \left[
    1+
    \chi r
    \frac{\gamma-1}{2}
    M_2^2
    \right],
    \label{eq:Tcav2}
\end{equation}
\noindent where $r=Pr^{1/3}$ is the recovery factor and $\chi$ is a recovery parameter. In the present implementation, $\chi=1$. The distinction between $T_2$ and $T_{\mathrm{cav},2}$ is important: $T_2$ is used as the lower physical bound of the effective-temperature model, while $T_{\mathrm{cav},2}$ is used to evaluate the modal-energy partition and vibrational relaxation.

\subsection{Modal-Energy Scales}
\label{sec:modal_energy_scales}

As Mach number and post-shock temperature increase, internal energy is distributed among translational, rotational, and vibrational modes. For simplicity, and because nitrogen is the dominant species in air, the following analysis uses nitrogen properties to construct the modal thermodynamic scales. The modal energies are evaluated using
\begin{equation}
    T_m=T_{\mathrm{cav},2}.
    \label{eq:modal_temperature_input}
\end{equation}
For a diatomic molecule,
\begin{equation}
    U_{\mathrm{trans}}
    =
    \frac{3}{2}R_uT_m,
    \qquad
    U_{\mathrm{rot}}
    =
    R_uT_m,
    \qquad
    U_{\mathrm{vib}}
    =
    \frac{
    R_u\theta_v
    }
    {
    \exp(\theta_v/T_m)-1
    },
    \label{eq:modal_energies}
\end{equation}
where $R_u$ is the universal gas constant. Zero-point vibrational energy is not included in the present formulation. The total modal energy is
\begin{equation}
    U_{\mathrm{tot}}
    =
    U_{\mathrm{trans}}
    +
    U_{\mathrm{rot}}
    +
    U_{\mathrm{vib}}.
    \label{eq:modal_total_energy}
\end{equation}

The characteristic vibrational temperature is obtained from
\begin{equation}
    \theta_v
    =
    \frac{
    hc\widetilde{\nu}_{N_2}
    }{k_B},
    \label{eq:theta_v}
\end{equation}
where the fundamental nitrogen vibrational wavenumber is
\begin{equation}
    \widetilde{\nu}_{N_2}
    =
    2358~\mathrm{cm^{-1}}.
    \label{eq:n2_wavenumber}
\end{equation}

To make it easier to estimate Rossiter modes, we introduce energy-weighted temperature scales that describe how the modal-energy distribution changes with the post-shock thermodynamic state. These are defined as
\begin{equation}
    T_{tt}
    =
    T_m
    \frac{
    U_{\mathrm{trans}}
    }{
    U_{\mathrm{tot}}
    },
    \qquad
    T_{rt}
    =
    T_m
    \frac{
    U_{\mathrm{rot}}
    }{
    U_{\mathrm{tot}}
    },
    \qquad
    T_{vt}
    =
    T_m
    \frac{
    U_{\mathrm{vib}}
    }{
    U_{\mathrm{tot}}
    }.
    \label{eq:weighted_modal_temperatures}
\end{equation}
From these quantities, two cumulative thermodynamic scales are constructed:
\begin{equation}
    T_{tr}
    =
    \left(
    T_{tt}T_{rt}
    \right)^{1/2},
    \label{eq:Ttr}
\end{equation}
and
\begin{equation}
    T_{trv}
    =
    \left(
    T_{tt}T_{rt}T_{vt}
    \right)^{1/3}.
    \label{eq:Ttrv}
\end{equation}

It is important to emphasize that $T_{tt}$, $T_{rt}$, $T_{vt}$, $T_{tr}$, and $T_{trv}$ are not literal nonequilibrium mode temperatures. Rather, they are surrogate thermodynamic scales constructed from the fraction of energy allocated to the translational, rotational, and vibrational degrees of freedom. The purpose of these quantities is to provide physically motivated bounds for an effective temperature that modifies the acoustic propagation timescale.

\subsection{Vibrational Relaxation and the Switching Mach Number}
\label{sec:relaxation_switch}

The remaining model choice is the location at which the effective-temperature representation transitions from a lower-Mach-number branch to a higher-Mach-number branch. Rather than prescribing this transition at a fixed Mach number, we relate it to the finite-rate vibrational response of the gas.

The post-shock pressure used in the relaxation calculation follows directly
from the normal-shock relation in Eq.~\eqref{eq:normal_shock_pressure},

\begin{equation}
    p_2
    =
    p_{\infty}
    \left[
    1+
    \frac{2\gamma}{\gamma+1}
    \left(
    M_{\infty}^2-1
    \right)
    \right].
    \label{eq:p2}
\end{equation}
The vibrational relaxation time is evaluated using the Millikan--White formulation for nitrogen collisions in air. For nitrogen colliding with species $j$,
\begin{equation}
    \tau_{N_2-j}
    =
    \frac{
    \exp
    \left[
    A_{N_2-j}
    \left(
    T_m^{-1/3}
    -
    B_{N_2-j}
    \right)
    -18.42
    \right]
    }
    {
    p_{2,\mathrm{atm}}
    },
    \label{eq:MW_tau}
\end{equation}
where $p_{2,\mathrm{atm}}$ is the post-shock pressure expressed in atmospheres. The Millikan--White coefficients are
\begin{equation}
    A_{N_2-j}
    =
    1.16\times10^{-3}
    \mu_{N_2-j}^{1/2}
    \theta_v^{4/3},
    \label{eq:MW_A}
\end{equation}
and
\begin{equation}
    B_{N_2-j}
    =
    0.015
    \mu_{N_2-j}^{1/4},
    \label{eq:MW_B}
\end{equation}
with reduced molecular weight
\begin{equation}
    \mu_{N_2-j}
    =
    \frac{
    W_{N_2}W_j
    }{
    W_{N_2}+W_j
    }.
    \label{eq:reduced_mass}
\end{equation}

For the present air approximation, the mixture-averaged vibrational relaxation time is
\begin{equation}
    \frac{1}{\tau_v}
    =
    \frac{
    X_{N_2}
    }{
    \tau_{N_2-N_2}
    }
    +
    \frac{
    X_{O_2}
    }{
    \tau_{N_2-O_2}
    },
    \label{eq:mixture_tau}
\end{equation}
where $X_{N_2}=0.79$ and $X_{O_2}=0.21$ are the mole fractions used in the relaxation model.

To compare the vibrational relaxation timescale with the characteristic cavity oscillation timescale, we evaluate the nondimensional parameter $\omega\tau_v$. Importantly, the oscillation frequency used for this diagnostic is evaluated independently of the effective-temperature branches so that the switching criterion is not circularly dependent on the model it selects. The post-shock sound speed is therefore taken as
\begin{equation}
    a_2
    =
    \sqrt{
    \gamma R_{\mathrm{mix}}T_2
    },
    \label{eq:a2}
\end{equation}
and the corresponding preliminary Rossiter estimate is
\begin{equation}
    St_{T_2}
    =
    \frac{
    n-\alpha
    }{
    \dfrac{1}{\kappa}
    +
    \dfrac{U_{\infty}}{a_2}
    }.
    \label{eq:St_T2}
\end{equation}
The dimensional frequency is then
\begin{equation}
    f_R
    =
    St_{T_2}
    \frac{
    U_{\infty}
    }{
    L
    },
    \label{eq:f_R_relax}
\end{equation}
and
\begin{equation}
    \omega\tau_v
    =
    2\pi f_R\tau_v.
    \label{eq:omega_tau}
\end{equation}

The present model does not define the transition by imposing $\omega\tau_v=1$. Instead, the physical switching Mach number is identified from the knee in the variation of $\log_{10}(\omega\tau_v)$ with Mach number. Defining
\begin{equation}
    y(M_{\infty})
    =
    \log_{10}
    \left(
    \omega\tau_v
    \right),
    \label{eq:log_omega_tau}
\end{equation}
the geometric curvature of this curve is
\begin{equation}
    \mathcal{K}(M_{\infty})
    =
    \frac{
    \left|
    \dfrac{d^2y}{dM_{\infty}^2}
    \right|
    }
    {
    \left[
    1+
    \left(
    \dfrac{dy}{dM_{\infty}}
    \right)^2
    \right]^{3/2}
    }.
    \label{eq:omega_tau_curvature}
\end{equation}
The switching Mach number is then defined as
\begin{equation}
    M_s
    =
    \underset{M_{\infty}}{\operatorname{arg\,max}}
    \;
    \mathcal{K}(M_{\infty}),
    \label{eq:M_switch}
\end{equation}
with a small guard region excluded near the ends of the Mach-number interval to prevent endpoint behavior of the numerical derivatives from selecting an artificial knee. Physically, this criterion identifies the Mach-number location at which the relaxation behavior changes most rapidly rather than assigning a transition based on a prescribed Mach number or an arbitrary modal-energy threshold.

\subsection{Physically Bounded Effective-Temperature Curves}
\label{sec:bounded_Teff}

The modal-energy analysis provides physical temperature scales that can be
used to constrain the effective temperature entering the acoustic-feedback
model. The initial formulation considered here constructs two power-law
branches in Mach number:
\begin{equation}
    T_{\mathrm{eff}}(M_{\infty})
    =
    \begin{cases}
        C_1M_{\infty}^{p_1},
        &
        M_{\infty}<M_s,
        \\[6pt]
        C_2M_{\infty}^{p_2},
        &
        M_{\infty}\geq M_s.
    \end{cases}
    \label{eq:Teff_two_curve}
\end{equation}

The coefficients $C_1$, $C_2$, $p_1$, and $p_2$ are not determined by
fitting the high-resolution numerical data. Instead, each branch is
constructed from physical boundary conditions obtained from the post-shock
and modal-energy calculations.

For the lower-Mach-number branch,
\begin{equation}
    T_{\mathrm{pre}}
    =
    C_1M_{\infty}^{p_1},
    \label{eq:Tpre}
\end{equation}
the left boundary is the normal-shock temperature at the lower end of the
modeled Mach-number interval,
\begin{equation}
    T_{\mathrm{pre}}(M_{\mathrm{low}})
    =
    T_2(M_{\mathrm{low}}),
    \label{eq:Tpre_left}
\end{equation}
while the right boundary is the translational--rotational scale at the
switching Mach number,
\begin{equation}
    T_{\mathrm{pre}}(M_s)
    =
    T_{tr}(M_s).
    \label{eq:Tpre_right}
\end{equation}
These two conditions give
\begin{equation}
    p_1
    =
    \frac{
    \ln
    \left[
    T_{tr}(M_s)/
    T_2(M_{\mathrm{low}})
    \right]
    }
    {
    \ln
    \left(
    M_s/M_{\mathrm{low}}
    \right)
    },
    \label{eq:p1}
\end{equation}
and
\begin{equation}
    C_1
    =
    \frac{
    T_2(M_{\mathrm{low}})
    }{
    M_{\mathrm{low}}^{p_1}
    }.
    \label{eq:C1}
\end{equation}

The initial higher-Mach-number branch was similarly represented using a
second power law,
\begin{equation}
    T_{\mathrm{post}}
    =
    C_2M_{\infty}^{p_2},
    \label{eq:Tpost}
\end{equation}
with the left boundary defined by the
translational-rotational-vibrational scale at the switching Mach number,
\begin{equation}
    T_{\mathrm{post}}(M_s)
    =
    T_{trv}(M_s),
    \label{eq:Tpost_left}
\end{equation}
and the right boundary defined by the same thermodynamic scale at a selected
upper Mach number,
\begin{equation}
    T_{\mathrm{post}}(M_{\mathrm{high}})
    =
    T_{trv}(M_{\mathrm{high}}).
    \label{eq:Tpost_right}
\end{equation}
The corresponding upper-branch coefficients are
\begin{equation}
    p_2
    =
    \frac{
    \ln
    \left[
    T_{trv}(M_{\mathrm{high}})
    /
    T_{trv}(M_s)
    \right]
    }
    {
    \ln
    \left(
    M_{\mathrm{high}}/M_s
    \right)
    },
    \label{eq:p2}
\end{equation}
and
\begin{equation}
    C_2
    =
    \frac{
    T_{trv}(M_s)
    }{
    M_s^{p_2}
    }.
    \label{eq:C2}
\end{equation}

Although both branches of this method are constrained by
physically motivated thermodynamic states rather than fitted directly to the
CFD discovered frequencies, an important limitation remains in the upper branch when using this first proposed method - the same limitation which leads to the need to define the second branch based on the temperature. In
particular, the exponent $p_2$ depends explicitly on the selected value of
$M_{\mathrm{high}}$. Unlike the switching Mach number $M_s$, which is
determined from the vibrational-relaxation response, $M_{\mathrm{high}}$ is
not itself determined by a physical transition criterion. It instead
represents the upper extent of the Mach-number interval selected for the
construction. The lower limit of the equation represented by $C_1$ and $p_1$ is not as limited, however from a practical use standpoint, the model's lower limit is Mach = 1, since below that we would assume that the acoustic propagation speed of sound does not change between the freestream and the cavity. 

As a result, changing the prescribed upper Mach number changes the slope of
$C_2M_{\infty}^{p_2}$ and therefore changes the predicted effective
temperature throughout the high-Mach-number branch. This dependence becomes
particularly undesirable when the model is applied outside the Mach-number
range originally used to construct the curve, since the prediction can vary
simply because a different value of $M_{\mathrm{high}}$ was selected.


To remove this dependence on an arbitrarily selected high-Mach-number bound,
the final formulation retains the lower-Mach-number power-law construction
but uses the translational--rotational--vibrational temperature scale
$T_{trv}$ directly after the relaxation-based switching point. The adopted
effective-temperature model is therefore
\begin{equation}
    T_{\mathrm{eff}}(M_{\infty})
    =
    \begin{cases}
        C_1M_{\infty}^{p_1},
        &
        M_{\infty}<M_s,
        \\[6pt]
        T_{trv}(M_{\infty}),
        &
        M_{\infty}\geq M_s.
    \end{cases}
    \label{eq:Teff_final}
\end{equation}

The lower branch remains unchanged and is therefore still determined
entirely by $T_2(M_{\mathrm{low}})$, $T_{tr}(M_s)$, and the
relaxation-based switching Mach number. Above $M_s$, however, no second
power-law exponent or upper-Mach-number boundary is required. Instead,
$T_{\mathrm{eff}}$ follows the $T_{trv}$ scale produced directly by the
modal-energy calculation.

This modification has two useful consequences. First, the high-Mach-number
behavior is determined only by the local thermodynamic state rather than by
the choice of an arbitrary $M_{\mathrm{high}}$. Second, because the
$T_{trv}$ scale is recalculated from the specified freestream conditions,
the high-Mach-number portion of the model is less sensitive to the Mach range
over which the model is evaluated. The resulting formulation is therefore
more suitable for comparisons between cases having different maximum Mach
numbers or different freestream thermodynamic conditions.

\begin{figure}
    \centering
    \includegraphics[width=1\linewidth]
    {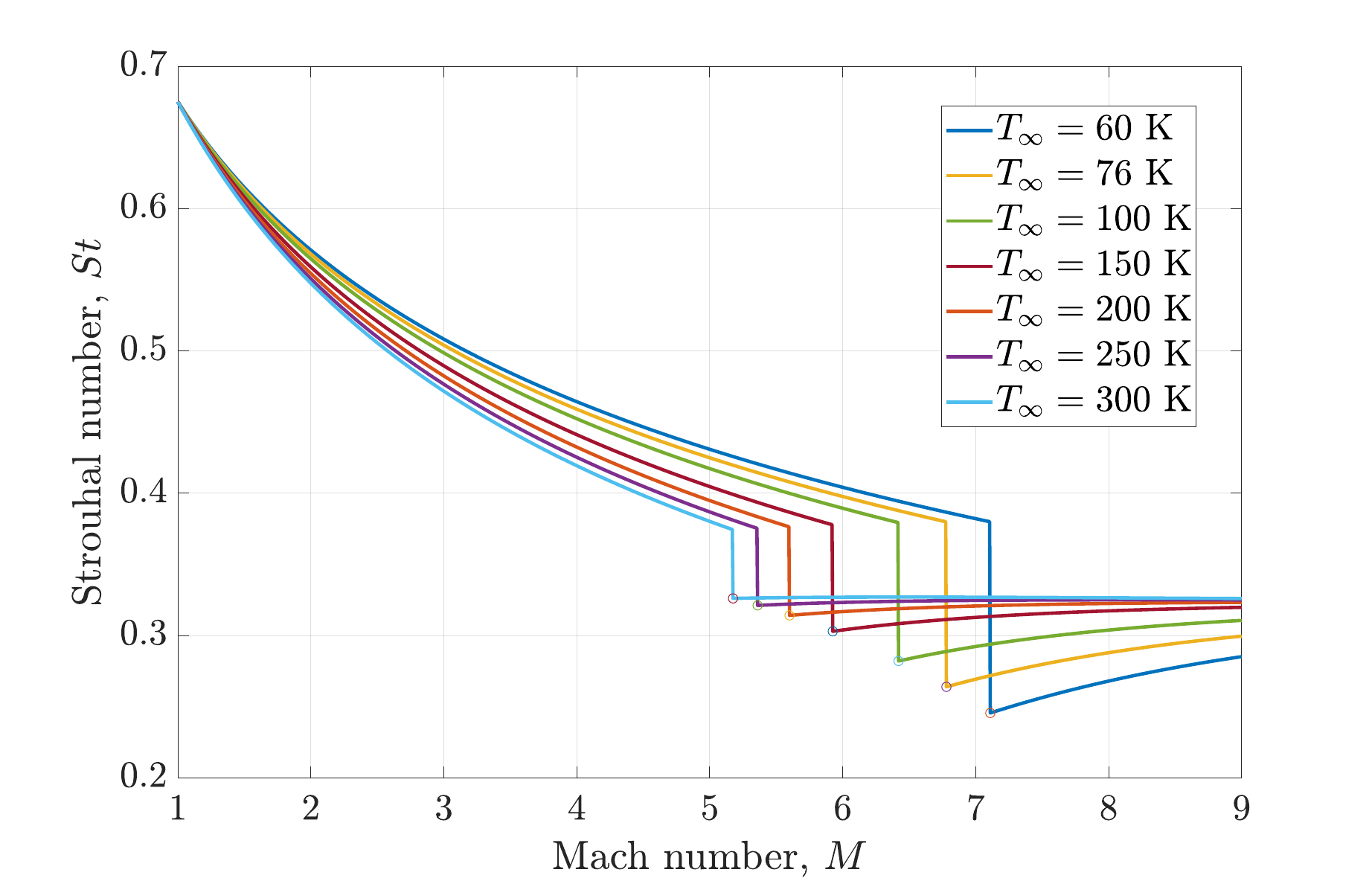}
    \caption{Final effective-temperature construction. The lower-Mach-number
    branch remains the physically bounded power law connecting
    $T_2(M_{\mathrm{low}})$ and $T_{tr}(M_s)$, while the
    translational--rotational--vibrational scale $T_{trv}$ is used directly
    for $M_{\infty}\geq M_s$. Unlike the two-power-law formulation in
    \autoref{eq:Teff_two_curve}, the high-Mach-number representation
    does not require the specification of an upper Mach-number boundary.}
    \label{fig:Teff_Ttrv}
\end{figure}

Figure~\ref{fig:Teff_Ttrv} shows the resulting construction when the upper
power-law branch is replaced by $T_{trv}$. In the cases examined here, the
direct $T_{trv}$ representation also provides good agreement with the
observed high-Mach-number frequency behavior while eliminating the need to
specify an upper Mach-number boundary.

It is again important to note that $T_{\mathrm{eff}}$, as well as the
modal-energy quantities used to construct it, does not represent a literal
nonequilibrium mode temperature. Instead, it is a surrogate thermodynamic
scale intended to reflect the Mach-number and temperature dependence of the
internal molecular degrees of freedom and their influence on the effective
acoustic response of the gas. The final construction therefore provides a
reduced-order representation of the changing thermodynamic response without
requiring the detailed cavity temperature field as an input and without
requiring an arbitrarily selected upper Mach number to define the
high-Mach-number behavior.

\subsection{Modified Rossiter Relation}
\label{sec:modified_rossiter}

The effective temperature is used to compute an effective acoustic propagation speed,
\begin{equation}
    a_{\mathrm{eff}}
    =
    \sqrt{
    \gamma R_{\mathrm{mix}}T_{\mathrm{eff}}
    },
    \label{eq:aeff}
\end{equation}
where $R_{\mathrm{mix}}$ is the specific gas constant of the mixture. This expression retains the simple perfect-gas form for the acoustic velocity, but the temperature entering the relation is now the physically bounded surrogate temperature described above.

The acoustic timescale is therefore
\begin{equation}
    \tau_a
    =
    \frac{L}{a_{\mathrm{eff}}},
    \label{eq:tau_a_eff}
\end{equation}
which can be substituted directly into Eq.~\eqref{eq:strouhalmod}. The final form of the proposed model is
\begin{equation}
    St_L
    =
    \frac{
    n-\alpha
    }{
    \dfrac{1}{\kappa}
    +
    \dfrac{U_{\infty}}{a_{\mathrm{eff}}}
    }.
    \label{eq:modified_rossiter}
\end{equation}

The DNS Strouhal-number data are used to evaluate the performance of
Eq.~\eqref{eq:modified_rossiter}, but they are not used to determine
$C_1$, $p_1$, or $M_s$. The lower effective-temperature branch is
constrained by the normal-shock and translational--rotational
thermodynamic scales, while the higher-Mach-number branch follows
$T_{trv}$ directly from the modal-energy calculation. The switching
Mach number $M_s$ is determined independently from the
vibrational-relaxation response.

Although $p_{\infty}$ does not appear explicitly in
Eq.~\eqref{eq:modified_rossiter}, the model is not independent of
freestream pressure. The freestream pressure sets the post-shock pressure
$p_2$, which directly affects the Millikan--White relaxation time. This
changes the shape of $\omega\tau_v$, the location of its knee, and
therefore the switching Mach number $M_s$. The switching Mach number in
turn affects the lower-branch coefficients $C_1$ and $p_1$ and determines
where the effective-temperature model transitions from the lower power-law
branch to the directly evaluated $T_{trv}$ branch. The pressure dependence
of the model can therefore be summarized as

\begin{equation}
\begin{aligned}
    p_{\infty}
    &\rightarrow
    p_2
    \rightarrow
    \tau_v
    \rightarrow
    \omega\tau_v
    \rightarrow
    M_s
    \\
    &\rightarrow
    \left(
    C_1,p_1,\text{branch selection}
    \right)
    \rightarrow
    T_{\mathrm{eff}}
    \rightarrow
    a_{\mathrm{eff}}
    \rightarrow
    St_L .
\end{aligned}
\label{eq:pressure_dependence}
\end{equation}

Similarly, changes in freestream temperature modify the post-shock state, the modal-energy scales, the vibrational-relaxation behavior, and the resulting switching location. Thus, the model does not assume universal Mach-number boundaries. Instead, its effective-temperature curves and switching Mach number are recomputed from the specified freestream thermodynamic state.

The resulting formulation preserves the basic Rossiter feedback structure while modifying the acoustic propagation timescale using a reduced-order thermodynamic model. This provides a first-order estimate of the cavity oscillation frequency without requiring prior knowledge of the detailed instantaneous thermodynamic conditions within the cavity.

\section{\label{sec:test}Model Evaluation}

To evaluate the model developed above, we use two dimensional numerical simulations based on the fully compressible form of Navier-Stokes equations with models to account for thermal and chemical nonequilibrium. These equations are solved using the open-source library, Eilmer \cite{gibbons2023eilmer}. Cases ranging from Mach 2 to 6 assumed ideal gas (i.e., no thermochemical nonequilibrium; results were identical when compared with simulations with non-equilibrum models switched on). For the cases at Mach 7 to 9, the 5 species non-equilibrium air model of \citet{gupta1990review} is used. For all cases, the initial composition of air in the freestream is assumed to be $N_2=0.767, O_2=0.233$. Time stepping is handled explicitly, with a third order Runge Kutta scheme. We use a fixed CFL number of 0.5, which leads to an average dt of $5.5\times10^{-9}$ s. 

\begin{figure}
    \centering
    \includegraphics[width=1\linewidth]{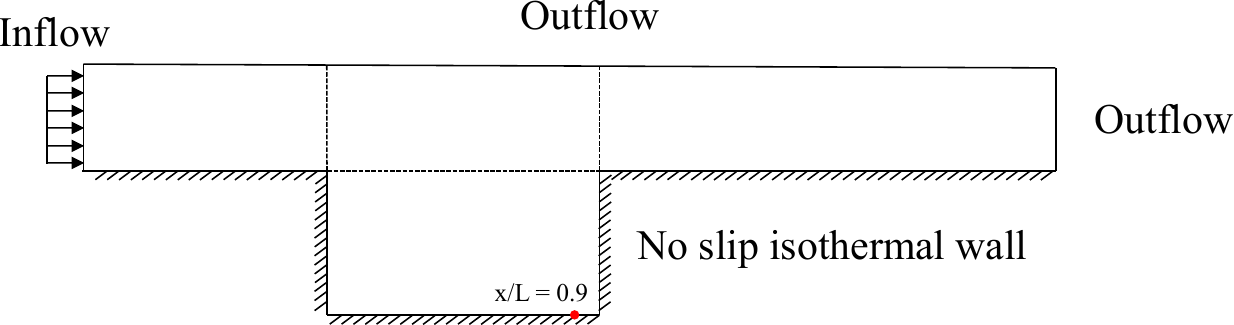}
    \caption{Schematic of the configuration. The pressure is measured at $x/L = 0.9$, similar to the study of \citet{rossiter1964wind}.}
    \label{fig:setup}
\end{figure}

Figure \ref{fig:setup} presents a schematic of the setup, indicating the boundary conditions and the probe location. A high refinement uniform grid is employed in both the x- and y-directions, meaning it is constructed to capture all relevant scales from the largest structures down to the Kolmogorov scale. As mentioned earlier, our simulations are two-dimensional; the absence of the vortex stretching term can moderately damp acoustic oscillations. Consequently, vortices are expected to survive longer in 2D computations. Additional information about the grid and boundary conditions is provided in our recent publications \citep{redding2025vortex, redding2023thermochemical}. It should also be noted that the grid used in this study was originally optimized for Reynolds numbers associated with Mach 11, so for the lower Mach numbers examined here, the resulting grid resolution is highly conservative. 
The initial flow conditions consists of $\mathbf{M_\infty} =
\begin{bmatrix}
2, 3, 4, 5, 6, 7, 8, 9
\end{bmatrix}, T_\infty = 76~\mathrm{K}$ and $p_\infty = 19,330~\mathrm{Pa}$. As shown in the figure, a probe is positioned on the cavity’s lower wall at $x/L=0.9$, identical to the location used by \citet{rossiter1964wind}. As mentioned before, in the present study, the dominant amplified mode corresponds to the second Rossiter mode; therefore, the mode number is taken as 2 in the initial numerical simulations from Mach 2 to 9, but multiple mode numbers are compared with experiment. 


\subsection{\label{sec:asymptote}Asymptotic limits}

The limits on the frozen speed of sound on the basis of chemical dissociation is then:

\begin{equation}
\overset{\text{lower bound}}{a_{f,L} = \sqrt{\frac{4}{3} R T}}
\;\leq\;
a_f
\;\leq\;
\overset{\text{upper bound}}{a_{f,U} = \sqrt{\frac{10}{3} R T}}
\label{eq:af_bounds}
\end{equation}
\\
and the equilibrium equation limits are:

\begin{equation}
\overset{\text{lower bound}}{a_{e,L} = \sqrt{\frac{4}{3}RT\tau}}
\;\leq\;
a_e
\;\leq\;
\overset{\text{upper bound}}{a_{e,U} = \sqrt{\frac{10}{3}RT\tau}}
\label{eq:ae_bounds}
\end{equation}

\begin{figure} 
    \centering
    \includegraphics[width=0.8\linewidth]{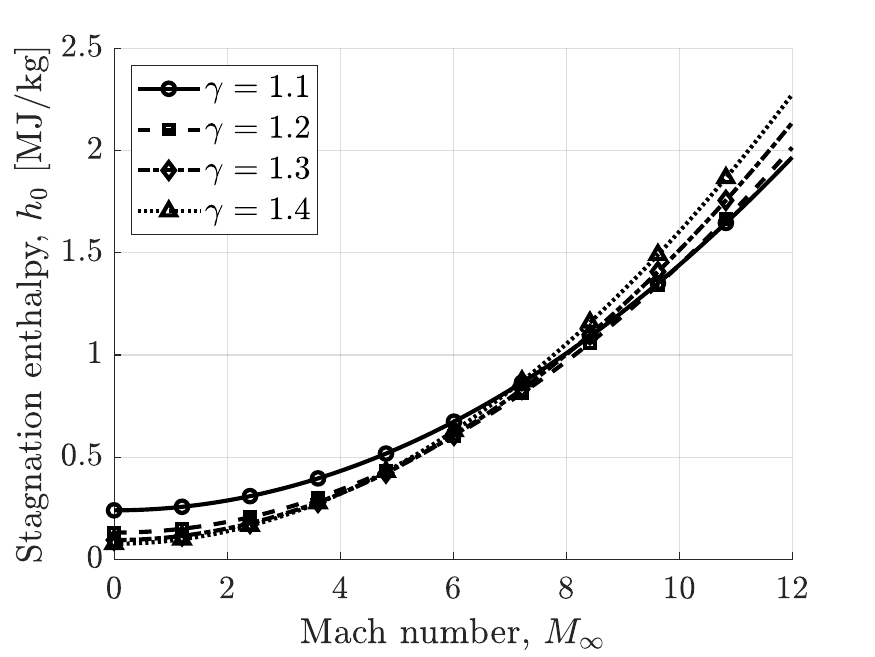}
    \caption{Comparison of Mach number versus stagnation enthalpy to show that enthalpy rises with Mach number in this configuration}
    \label{fig:enthalpy}
\end{figure}

Substituting these limiting functions into the Strouhal equation in place of the usual freestream speed of sound, i.e., replacing $\tau_a$ with $L/a_{e,f}$ provides the asymptotes for the Strouhal number. \autoref{tab:strouhal_bounds} lists the comparison between simulation results with frozen limits and the Heller-Bliss estimates. Before showing this, it is important to note that Mach number is used instead of enthalpy in the x-axis of our plots and figures, because it is useful as a comparative parameter. However, since flow stagnation enthalpy can have a significant effect on this flow phenomena, it is also critical to show how flow stagnation enthalpy changes with Mach number, to illustrate that in this configuration, if the flow Mach number increases, the flow stagnation enthalpy increases as well. $\gamma$ has an effect on the slope of the relation, but the correlation is largely linear. Since the goal of this model is to provide the researcher with a method to compute Strouhal number from solely apriori knowledge of Mach number, this is a helpful comparison. As the cavity flow experiences further reactions, the effective value of gamma will change in the field. As additional thermochemical reactions or internal energy modes become active, the effective specific heat increases and as a result the ratio of specific heats $\gamma$ goes down. A range of gamma values is plotted to show how significantly this change effects the correlation between Mach and enthalpy in \autoref{fig:enthalpy}.

\begin{table}
\centering
\caption{Comparison of Strouhal number from present DNS calculations with those of \citet{heller1975physical} model. Also included are the chemically frozen limits obtained from equation \ref{eq:af_bounds}. Red font indicates values out of the frozen bounds, and orange is nearly outside the bounds. Note, Mach 7 and above utilized non-equilibrium gas models in CFD.}
\label{tab:strouhal_bounds}
\begin{tabular}{lcccc}
\toprule
\textbf{Mach} & \textbf{$St_{f,L}$} & \textbf{$St_{\mathrm{CFD}}$} & \textbf{$St_{\mathrm{Model}}$}\citep{heller1975physical} & \textbf{$St_{f,U}$} \\
\midrule
2 & 0.484 & 0.56 & 0.57 & 0.696 \\
4 & 0.40  & 0.45 & \textbf{\textcolor{orange}{0.49}} & 0.49 \\
6 & 0.23  & 0.38 & \textbf{\textcolor{red}{0.478}} & 0.41 \\
7 & 0.203 & 0.28 & \textbf{\textcolor{red}{0.474}} & 0.369 \\
8 & 0.18  & 0.32 & \textbf{\textcolor{red}{0.471}} & 0.34 \\
9 & 0.16  & 0.29 & \textbf{\textcolor{red}{0.469}} & 0.31 \\
\bottomrule
\end{tabular}
\end{table}

The effective acoustic propagation speed is bounded by frozen and thermochemical equilibrium limits. Frozen conditions represent minimal internal energy exchange, while equilibrium conditions represent full thermochemical relaxation. Hypersonic flows generally exist between these limits due to finite-rate vibrational and rotational energy exchange. The effective sound speed therefore represents a reduced-order description capturing intermediate relaxation behavior rather than a purely empirical curve fit. A few numerical observations can be made from \autoref{tab:strouhal_bounds}. First, the Strouhal number at Mach 2 from numerical simulations agree well with \citet{heller1975physical} model as this model is validated up to Mach 3. Second, as expected, there is a decreasing trend in Strouhal number as Mach number increases. This is where the similarities between simulations and previous models end. Given that previous models assumed the gas to be non-reactive and calorically perfect, the Strouhal number should lie within the bounds of frozen and equilibrium flows. However, as the Mach number increases beyond 4, the \citet{heller1975physical} estimate diverges from these bounds. On the other hand, our simulations, especially those up to Mach 6 where the air was assumed to be ideal, indicate that while $St$ decreases, it stays in the aforementioned bounds. Further, to confirm our hypothesis, we conducted calculations for Mach numbers up to 6 while modeling nitrogen dissociation and found that the results did not change. This makes sense because at these Mach numbers, we do not expect much dissociation in the given configuration.

\begin{figure}
    \centering
    \includegraphics[width=1.0\linewidth]{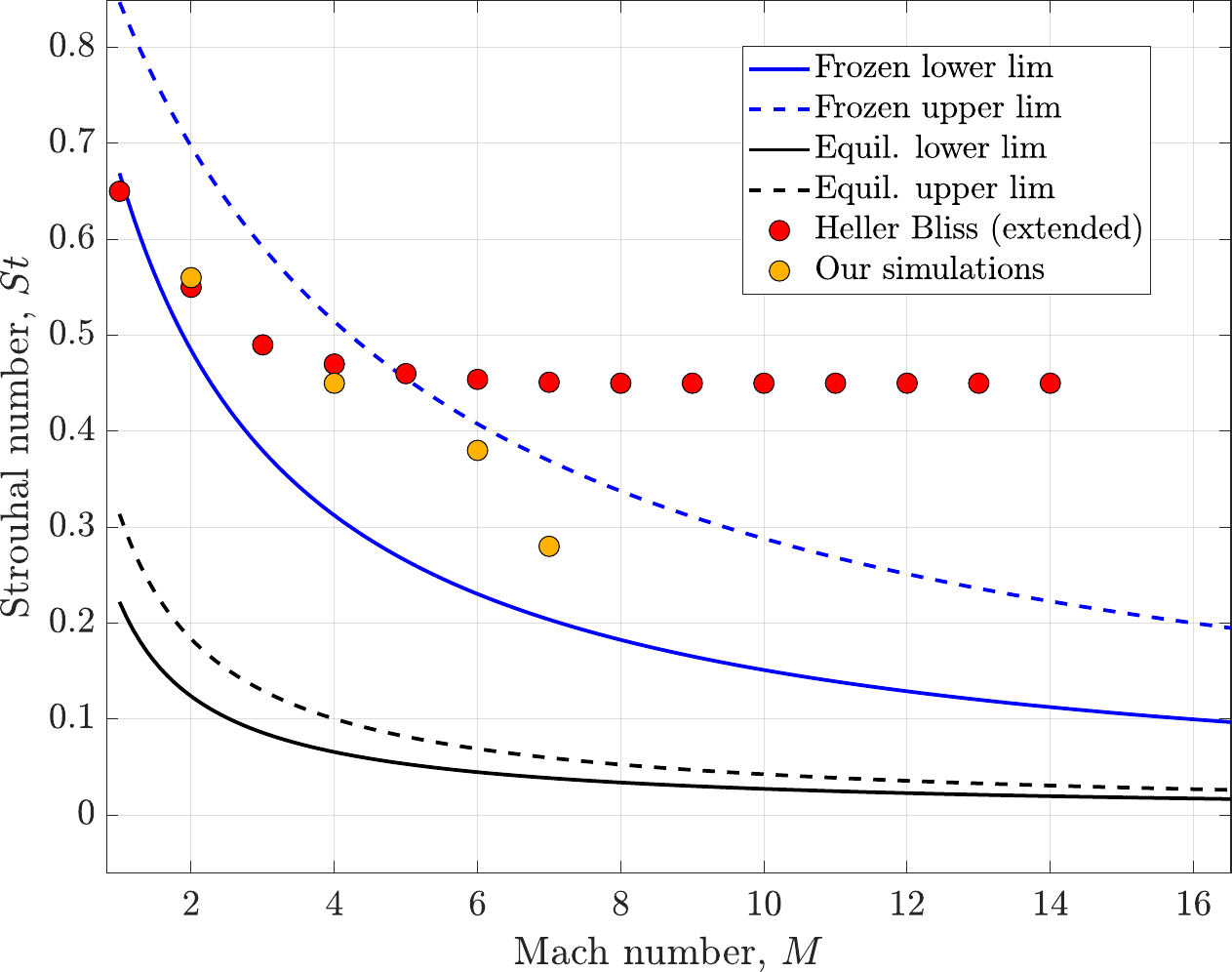}
    \caption{Mach number verses Strouhal number for present freestream temperature, extended to Mach 16, all models generated assuming second mode excitation (n=2), which carries the most energetic contribtution in the DNS}
    \label{fig:settempmachvstrouhal}
\end{figure}

To investigate these limits further, we plot the frozen and equilibrium assumptions for our current freestream temperature verses Mach number. As shown in \autoref{fig:settempmachvstrouhal}, as Mach number increases, the Strouhal number function begins to approach a limit. Again, the equilibrium assumption approaches its asymptote much more quickly than that of the frozen assumption. It is important here to note the significant divergence between the Heller-Bliss model and the one predicted by employing a speed of sound found by the frozen assumption. While the Heller-Bliss asymptote is $St=0.45$, the frozen flow limit does not occur until much higher in Mach number, and appears to be between 0.08 for the lower frozen limit and 0.15 for the upper frozen limit.  

\subsection{\label{sec:finite}Evaluation of theoretical model at finite Mach numbers}
 
for which the freestream temperature is $T_{\infty}=76$ K and the freestream
Mach number is varied from 2 to 9. The remaining freestream and geometric
parameters are held at the values defined previously for the baseline
configuration. For each Mach number, the effective acoustic state is determined
using the curves described in
Section~\ref{sec:theory}. The transition between the branches is determined
from the knee in $\log_{10}(\omega\tau_v)$ and is therefore calculated from
the thermodynamic and relaxation state rather than prescribed at a fixed Mach
number or fit to the DNS frequency data.

\begin{figure}
    \centering
    \includegraphics[width=1.0\linewidth]{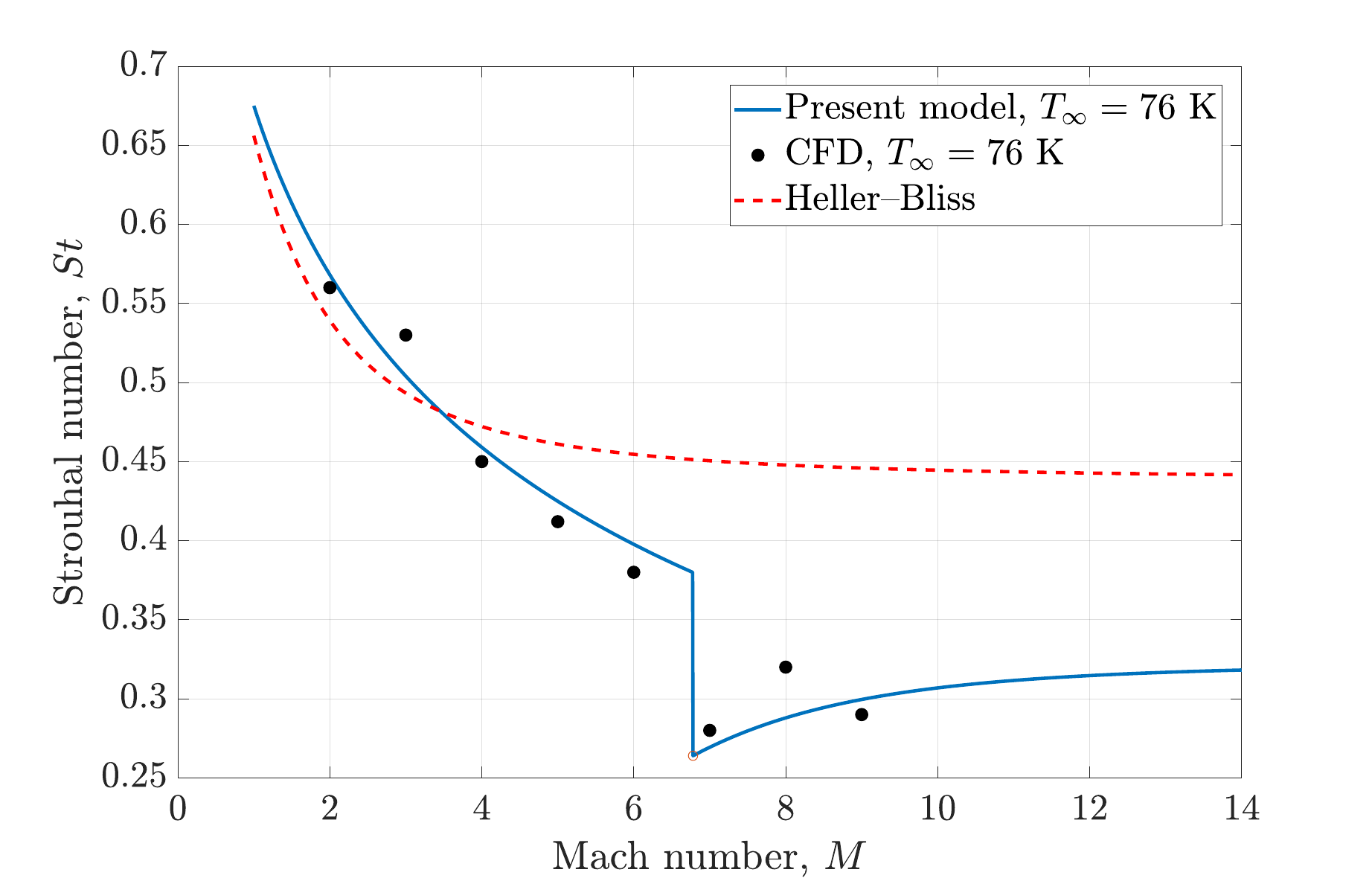}
    \caption{Strouhal number as a function of freestream Mach number for the
    present $T_{\infty}=76$ K configuration. DNS results are compared with
    the proposed thermodynamically modified Rossiter model and the
    Heller-Bliss model \citep{heller1975physical}. The proposed model uses
    two physically bounded effective-temperature branches separated by the
    computed relaxation-based switching Mach number. Results shown correspond
    to the second Rossiter mode, $n=2$.}
    \label{fig:strouhalregimes}
\end{figure}

Figure~\ref{fig:strouhalregimes} shows the resulting Strouhal-number
predictions together with the values obtained from the DNS calculations.
The Heller--Bliss model \citep{heller1975physical} is also shown for
comparison. At the lower Mach numbers, both models provide similar
predictions. As Mach number increases, however, the Heller--Bliss prediction
approaches an approximately constant Strouhal number, whereas the DNS exhibits
a substantial reduction in Strouhal number. The proposed model captures this
decrease through Mach 7 and also reproduces the subsequent recovery in
Strouhal number at the highest Mach numbers considered. Although the precise
non-monotonic variation of the DNS is not reproduced at every Mach number,
the thermodynamically modified formulation follows the overall high-Mach-number
behavior considerably more closely than the conventional compressibility
correction.

Importantly, the change in the predicted behavior at higher Mach number is not
introduced through a prescribed Mach-number cutoff. The transition follows
from the finite-rate vibrational-relaxation response used to determine the
switching location, while the two effective-temperature curves are constrained
by the post-shock and modal-energy states described in
Section~\ref{sec:theory}. Thus, neither the switching Mach number nor the
effective-temperature curves are obtained by fitting the DNS Strouhal-number
data.

To further quantify the accuracy of the model,
Table~\ref{tab:cfd_vs_analytical} compares the predicted Strouhal numbers
with the DNS values. Absolute percentage error is defined relative to the DNS
result as
\begin{equation}
    \epsilon
    =
    100
    \left|
    \frac{St_{\mathrm{pred}}-St_{\mathrm{DNS}}}
    {St_{\mathrm{DNS}}}
    \right|.
    \label{eq:percent_error}
\end{equation}
The proposed model remains within approximately 10.3\% of the DNS values
throughout the Mach-2-to-9 range, with the largest discrepancy occurring at
Mach 8. At Mach 6, for example, the predicted Strouhal number is 0.384
compared with the DNS value of 0.38. The difference between the two models
becomes substantially larger at higher Mach number. Using the same error
definition, the Heller-Bliss prediction differs from the numerical simulations by approximately
69\% at Mach 7 and 62\% at Mach 9, whereas the proposed model differs by
approximately 5.7\% and 3.1\%, respectively.

\begin{table}[t]
\centering
\caption{Comparison of CFD-estimated Strouhal number at $x/L=0.9$ with
the proposed model and Heller-Bliss (HB) for the present
$T_{\infty}=76$ K configuration. Errors are absolute percentage differences
relative to the simulation result.}
\label{tab:cfd_vs_analytical}

\begin{tabular}{cccccc}
\toprule
\textbf{$M_\infty$}
& \textbf{$St_{\mathrm{CFD}}$}
& \textbf{$St_{\mathrm{model}}$}
& \textbf{Error (\%)}
& \textbf{$St_{\mathrm{HB}}$}
& \textbf{HB Error (\%)} \\
\midrule

2
& 0.560
& 0.568
& 1.46
& 0.539
& 3.70 \\

3
& 0.530
& 0.504
& 4.90
& 0.493
& 6.92 \\

4
& 0.450
& 0.459
& 2.01
& 0.472
& 4.93 \\

5
& 0.412
& 0.425
& 3.13
& 0.461
& 11.91 \\

6
& 0.380
& 0.398
& 4.65
& 0.455
& 19.63 \\

7
& 0.280
& 0.269
& 3.82
& 0.451
& 60.90 \\

8
& 0.320
& 0.288
& 10.03
& 0.448
& 39.94 \\

9
& 0.290
& 0.300
& 3.29
& 0.446
& 53.77 \\

\bottomrule
\end{tabular}
\end{table}

During the finite-rate chemistry simulations, under the present conditions
only trace concentrations of product species are generated and the mixture
remains close to its freestream composition. Consequently, variations in
mixture thermodynamic properties such as $c_p$, $\gamma$, and composition
across the cases remain comparatively small. The increase in total flow
enthalpy is instead dominated by the kinetic-energy contribution associated
with increasing Mach number. Mach number is therefore retained as the primary
organizing parameter because it provides a compact and conventional measure of
the energetic state of the flow while maintaining direct comparability with
previous cavity-flow studies.

The comparison with the present DNS provides an internally consistent
evaluation of the model over a broad Mach-number range. However, agreement
with simulations performed using a single baseline configuration is not by
itself sufficient to establish whether the model can be applied to different
freestream temperatures, cavity dimensions, or experimental facilities.
The model is therefore additionally evaluated using independent numerical
and experimental cavity-flow cases that were not used in its development.
These comparisons are summarized in Table~\ref{tab:model_validation}.

One additional numerical case was performed using the present cavity
configuration at $M_\infty=6$, $T_\infty=216$ K, and
$p_\infty=19661$ Pa. The DNS gives $St=0.31$, while the proposed model
predicts $St=0.33$, corresponding to an absolute difference of 6.45\%.
Because this case does not have a corresponding experimental measurement,
it is included as an additional numerical evaluation but is not assigned an
experimental error in Table~\ref{tab:model_validation}.

Independent experimental comparisons provide a more stringent test of the
model. In the Mach-6.14 experiment of \citet{Benitez_Bisek_Borg_Clingenpeel_Molinaro_Meritt_2026a}, for which
$T_\infty=52.1$ K and $U_\infty=888.5$ m/s, the measured peak
frequency is approximately 5000 Hz. The present model predicts 4989.2 Hz at mode 4, which is lower than the Rossiter baseline prediction of this mode occurring around mode 7. This corresponds to an error of 0.2\%.

The Mach-4.95 experiments of \citet{unalmis2001experimental} provide a
particularly useful comparison because several cavity length-to-depth ratios
and multiple oscillation modes were measured under the same freestream
conditions. At $T_\infty=59.8$ K and $U_\infty=765$ m/s, the present model
predicts the second-mode frequencies to within approximately 0.5--3.2\% for
$L/D=3$, 4, and 5. For $L/D=6$, the predicted first- and second-mode
frequencies both fall within the experimental frequency ranges. The first-mode
predictions at $L/D=3$ and 4 differ from the measurements by approximately
10.0\% and 13.3\%, respectively.

These Mach-4.95 measurements are also relevant to the physical interpretation
of the present model. \citet{unalmis2001experimental} found that the measured
frequencies at Mach 5 approached the longitudinal acoustic modes of a closed
cavity and concluded that shear-layer/acoustic coupling was substantially
weaker than in lower-Mach-number cavity flows. The ability of the present
model to reproduce these frequencies while retaining the convective and
phase-delay components of the Rossiter feedback relation supports the use of
a thermodynamically modified acoustic timescale rather than replacing the
feedback model entirely with a closed-box acoustic representation.

A further Mach-5 experimental comparison is made with the measurements of
Diaz Villa, Sirohi, and Clemens \citep{diazvilla2025store}. Their experiment
used $T_\infty=57.5$ K, $U_\infty=760$ m/s, and a cavity length of
$L=158.75$ mm. A measured frequency of 980 Hz is compared with the present
first-mode prediction of 918 Hz, corresponding to an absolute difference of
6.33\%. A higher measured peak at approximately 3200 Hz is compared with the
present third-mode prediction of 3212 Hz, resulting in an absolute difference
of only 0.38\%. The latter comparison uses the modal numbering associated with
the present Rossiter formulation.

Finally, the lower-Mach-number behavior of the model is assessed using a
separate Mach-2 experimental data set from Ünalmiş et al. For
$M_\infty=2$, $T_\infty=155.2$ K, $U_\infty=510$ m/s, and
$L=0.1524$ m, the measured first- and second-mode Strouhal numbers are
0.24 and 0.56. The proposed model predicts 0.23 and 0.53, corresponding to
absolute errors of 4.17\% and 5.36\%, respectively.

\begin{landscape}

\begin{table*}[t]
\centering
\footnotesize
\caption{Comparison of the present model with additional numerical and
experimental cavity-frequency data. Experimental error is calculated as
$100\times\left|(\mathrm{Model}-\mathrm{Experiment})/
\mathrm{Experiment}\right|$. The error entry is left blank for the
CFD-only comparison because no corresponding experimental measurement is
available.}
\label{tab:model_validation}

\begin{tabular}{lcccccccccc}
\toprule
\textbf{Source}
& \textbf{$M_\infty$}
& \textbf{$T_\infty$}
& \textbf{$p_\infty$}
& \textbf{$U_\infty$}
& \textbf{Geometry}
& \textbf{Mode/}
& \textbf{Experiment}
& \textbf{CFD}
& \textbf{Model}
& \textbf{Error } \\
& & (K) & (Pa)& (m/s) & & \textbf{Quantity}& & & & (\%)\\
\midrule

Present CFD
& $6.00$ & $216$ & $19661$ &
& Present cavity
& $St$
& --
& 0.310
& 0.330
& \\

\citet{Benitez_Bisek_Borg_Clingenpeel_Molinaro_Meritt_2026a}
& $6.14$ & $52.1$ & & $888.5$ 
& --
& $f$
& ~5000 Hz 
& --
& 4989.2 Hz 
& 0.2 \\
 &&&&&&&&&(@ $mode =4$)& \\
\citet{unalmis2001experimental}
& $4.95$ & $59.8$ & & $765$ 
& $L/D=3$
& $f_1$
& 2.11 kHz
& --
& 1.90 kHz
& 9.95 \\

\citet{unalmis2001experimental}
& $4.95$& $59.8$ & &  $765$ 
& $L/D=3$
& $f_2$
& 4.32 kHz
& --
& 4.30 kHz
& 0.46 \\

\citet{unalmis2001experimental}
& $4.95$& $59.8$ & &  $765$ 
& $L/D=4$
& $f_1$
& 1.66 kHz
& --
& 1.44 kHz
& 13.25 \\

\citet{unalmis2001experimental}
& $4.95$& $59.8$ & &  $765$ 
& $L/D=4$
& $f_2$
& 3.16 kHz
& --
& 3.26 kHz
& 3.16 \\

\citet{unalmis2001experimental}
& $4.95$& $59.8$ & &  $765$ 
& $L/D=5$
& $f_2$
& 2.53 kHz
& --
& 2.61 kHz
& 3.16 \\

\citet{unalmis2001experimental}
& $4.95$& $59.8$ & &  $765$ 
& $L/D=6$
& $f_1$
& 0.85--1.15 kHz
& --
& 0.969 kHz
& Within range \\

\citet{unalmis2001experimental}
& $4.95$& $59.8$ & &  $765$ 
& $L/D=6$
& $f_2$
& 2.06--2.26 kHz
& --
& 2.15 kHz
& Within range \\

\citet{diazvilla2025store}
& $5.00$& $57.5$ & &  $760$ 
& $L=158.75$ mm
& $f_1$
& 980 Hz
& --
& 918 Hz
& 6.33 \\

\citet{diazvilla2025store}
& $5.00$& $57.5$ & &  $760$ 
& $L=158.75$ mm
& $f_3$
& 3200 Hz
& --
& 3212 Hz
& 0.38 \\

\citet{unalmis2001experimental}
& $2.00$& $155.2$ & &  $510$ 
& $L=0.1524$ m
& $St_1$
& 0.240
& --
& 0.230
& 4.17 \\

\citet{unalmis2001experimental}
& $2.00$& $155.2$ & &  $510$ 
& $L=0.1524$ m
& $St_2$
& 0.560
& --
& 0.530
& 5.36 \\

\bottomrule

\multicolumn{11}{l}{The present Mach-6 CFD case has no corresponding experimental measurement;
the proposed model differs from the CFD Strouhal} \\
\multicolumn{11}{l}{ number by 6.45\%.
For the $L/D=6$ \citet{unalmis2001experimental} cases, the model predictions fall within the reported experimental frequency ranges.}
\end{tabular}%


\end{table*}

\end{landscape}
Taken together, the comparisons in Tables~\ref{tab:cfd_vs_analytical}
and~\ref{tab:model_validation} demonstrate the behavior of the proposed
formulation over substantially different freestream temperatures, Mach
numbers, cavity dimensions, and oscillation modes. The model retains reasonable
agreement with the lower-Mach-number measurements, where the classical
Rossiter feedback mechanism remains appropriate, while also reproducing the
frequency behavior of higher-Mach-number cases in which the cavity response
has been observed to become increasingly acoustic.

The present closure is not intended to serve as a universal thermochemical
model. Rather, it is a reduced-order representation of the changing effective
acoustic propagation speed that retains the convective and phase-delay
components of the Rossiter feedback framework. The agreement with both the
present DNS and independent experimental measurements provides support for
using this thermodynamically modified acoustic timescale to extend the
predictive range of the classical Rossiter formulation.

\section{\label{sec:conclusion}Conclusion}

A new model for computing Strouhal number based on freestream temperatures is presented. The Heller-Bliss model is shown to diverge from the CFD results, while the adapted model shows close alignment with the CFD results, and is physically reasoned on the basis of energy modes. Two methods are shown with increasing precision. As a first prediction approach, one could calculate the minimum and maximum speed of sound, and utilize this new value in the Strouhal number calculation to predict a minimum and maximum Strouhal value. This is a more rapid approach to determine ``problem frequency ranges'' of interest for the specific problem, and whether or not a finer approach is needed. Separately, for a better prediction, the second method can be employed. This method has two distinct regimes on the basis of temperatures and energy modes. Using these temperatures in the speed of sound calculations, the Rossiter equation yields closer predictions to the exact Strouhal number.  
\section*{Acknowledgements}
Research was sponsored by the DEVCOM Army Research Laboratory and was accomplished in-part under Cooperative Agreement Number W911NF-24-2-0152 and W911NF-23-2-0026. Luis Bravo was supported by the 6.1 program in weapon propulsion sciences. The authors gratefully acknowledge the High-Performance Computing Modernization Program (HPCMP) resources and support provided by the Department of Defense Supercomputing Resource Center (DSRC) as part of the 2022 Frontier Project, Large-Scale Integrated Simulations of Transient Aerothermodynamics in Gas Turbine Engines.

The views and conclusions contained in this document are those of the authors and should not be interpreted as representing the official policies, either expressed or implied, of the DEVCOM Army Research Laboratory or the U.S. Government. The U.S. Government is authorized to reproduce and distribute reprints for Government purposes notwithstanding any copyright notation herein.


\section*{ORCID iDs}

Jeremy Redding: \href{https://orcid.org/0000-0001-8012-2101}{0000-0001-8012-2101}

Luis Bravo: \href{https://orcid.org/0000-0003-1604-9303}{0000-0003-1604-9303}

Prashant Khare: \href{https://orcid.org/0000-0002-5810-4377}{0000-0002-5810-4377}

\section*{Declaration of Interests}

The authors report no conflict of interest to disclose. 

\nocite{*}

\bibliography{jfm.bib}

\end{document}